\documentclass[10pt]{article}
\usepackage{amsmath,amssymb,amsthm}
\usepackage{graphicx}
\newtheorem{theorem}{Theorem}
\newtheorem{definition}{Definition}
\newtheorem{corollary}{Corollary}
\newtheorem{proposition}{Proposition}
\newtheorem{axiom}{Axiom}
\newtheorem{example}{Example}

\def\orthrel{\widehat{\perp}}
\def\zero{\bar{\mathtt{0}}}
\def\one{\bar{\mathtt{1}}}

\title{State Property Systems and Orthogonality}
\author{Diederik Aerts and Didier Deses \footnote{Research Assistant of the Fund for Scientific Research
Flanders (Belgium)}}
\date{}

\begin{document}

\maketitle

\scriptsize
\centerline{Foundations of the Exact Sciences (FUND) and Topology (TOPO)}
\centerline{Department of Mathematics, Brussels Free University, Pleinlaan 2, 1050 Brussels.}
\centerline{{\tt diraerts@vub.ac.be, diddesen@vub.ac.be}}
\normalsize

\begin{abstract}
\noindent The structure of a state property system was introduced to formalize in a complete way the operational
content of the Geneva-Brussels approach to the foundations of quantum
mechanics \cite{Aerts1999a,Aerts1999b,ACVVVS1999}, and the category of state property systems was proven to be
equivalence to the category of closure spaces \cite{ACVVVS1999,ACVVVS2001}. The first
axioms of standard quantum axiomatics (state determination and atomisticity)
have been shown to be equivalent to the $T_0$ and $T_1$ axioms of closure
spaces \cite{VanSteirteghem2000, VanderVoorde2000,VanderVoorde2001}, and
classical properties to correspond to clopen sets, leading to a decomposition
theorem into classical and purely nonclassical components for a general state
property system \cite{ADVV2001,ADVV2002,AD2002}. The concept of orthogonality,
very important for quantum axiomatics, had however not yet been introduced within the formal scheme of the state
property system. In this paper we introduce orthogonality in a operational way, and define ortho state property
systems. Birkhoff's well known biorthogonal construction gives rise to an orthoclosure and we study the relation
between this orthoclosure and the operational orthogonality that we introduced.
\end{abstract}

\section{Introduction}
Within the Geneva-Brussels approach to the Foundations of Quantum Mechanics
\cite{Piron1976,Aerts1981,Aerts1982,Aerts1983,Piron1989,Piron1990} the basic operational concept to
construct the theory is that of a `test' (in some articles also called `yes/no experiment', `question'
or `operational project'). For a physical entity $S$ one considers the set of all relevant tests $Q$,
and denotes tests by means of symbols $\alpha,
\beta, \gamma, \ldots \in Q$. The basic ontological concept is that of state of the physical entity $S$,
and the set of all relevant states is denoted by $\Sigma$, while individual states are denoted by symbols
$p, q, r, \ldots \in \Sigma$. A basic structural law on $Q$ is the following: ``if the
entity $S$ is in a state $p \in \Sigma$, such that the outcome `yes' is certain for
$\alpha$, then the outcome `yes' is certain for $\beta$". If this is satisfied we say that $\alpha$
implies $\beta$ and denote $\alpha < \beta$. This law defines a pre-order relation on $Q$, which
induces an equivalence relation on $Q$. A property of the entity $S$ is then introduced as the
equivalence class of tests that test this property, the set of all relevant properties is denoted by
${\cal L}$ and individual properties by symbols $a, b, c, \dots \in {\cal L}$.
Two operations are introduced operationally on $Q$. For an arbitrary test $\alpha \in Q$
the `inverse test' $\widetilde\alpha$ is introduced, which is the test that
consists of performing $\alpha$ and exchanging the role of `yes' and `no', and it is demanded that
$\widetilde{\ }: Q \rightarrow Q$ is defined on all $Q$, and obviously
$\widetilde{\widetilde\alpha}= \alpha$. For an arbitrary collection $\{\alpha_i\}$ of tests the `product
test' $\Pi_i\alpha_i$ is defined as the test that consists of choosing one of
the $\alpha_i$ and performing the chosen test and interpreting the outcome thus obtained as outcome of
$\Pi_i\alpha_i$. It is also demanded that $Q$ is closed for the product operation on tests, and as a
consequence it can be proven that the set of properties ${\cal L}$ is a complete
lattice, for the trace on ${\cal L}$ of the pre-order relation on $Q$, which is a `partial
order relation' on ${\cal L}$, denoted $<$, with the meaning: $a < b$ iff whenever the state of the
entity $S$ is such that $a$ is actual, then also $b$ is actual. The infimum for a collection of
properties $\{a_i\}$, $a_i \in {\cal L}$, is denoted $\wedge_ia_i$ and it is the equivalence
class of the product test $\Pi_i\alpha_i$, where for each $i$ the test $\alpha_i$ tests the property
$a_i$, hence the physical meaning of the infimum property is the conjunction.

More recently almost (and we come to this immediately) the whole scheme that is
obtained as such in a purely operational way was formalized by introducing the
structure of a state property system \cite{Aerts1999a,Aerts1999b,ACVVVS1999}.

\begin{definition}[State Property System]
\label{def:statprop} A triple $(\Sigma,{\cal L},\xi)$ is called a
state property system if $\Sigma$ is a set, ${\cal L}$ is a
complete lattice and $\xi :  \Sigma \rightarrow {\cal P}({\cal
L})$ is a function such that for $p \in \Sigma$, $\zero$ the minimal
element of ${\cal L}$ and $(a_i)_i \in {\cal L}$, we have
$\zero\not\in\xi(p)$ (SPS1) and $a_i\in \xi(P),\forall i$ implies
$\wedge_ia_i\in\xi(p)$ (SPS2). Moreover for $a, b\in {\cal L}$ we have
that $a < b$ if and only if for every $r \in \Sigma$:$a \in \xi(r)$ implies
$b \in \xi(r)$ (SPS3).
\end{definition}
\noindent
It is by the introduction of the function $\xi$ that the state property system formalizes
the operational content of the Geneva-Brussels approach. The physical meaning of $\xi(p)$ for an
arbitrary state $p \in \Sigma$ of the physical entity $S$, is that $\xi(p)$ is the set of all
properties that are actual when $S$ is in state $p$. This makes it clear why (SPS1),
(SPS2) and (SPS3) have to be satisfied. Indeed,
(SPS1) expresses that $0$, the minimal property, is the property that is never actual,
for example the property `this entity $S$ is not there'. And (SPS2) expresses that the
infimum of properties that are actual in a state is also an actual property, which has to be so because
of the physical meaning of conjunction for the infimum. And (SPS3) expresses the
physical law: $a < b$ iff whenever the state of the
entity $S$ is such that $a$ is actual, then also $b$ is actual.

We mentioned already that the state property system only manages to capture `almost' all of the
operational structure. Indeed the structure of the inverse operation $\widetilde{\ }: Q \rightarrow Q$,
was not captured within the formal structure of the state property system. The reason why there is
a fundamental problem here is because the inverse on the set of tests
does not transpose to an operation on the set of properties by means of the quotient. This is because for
two equivalent tests $\alpha, \beta \in Q$ we do not in general have that the inverse tests
$\widetilde\alpha$ and
$\widetilde\beta$ are equivalent. The problem was known in the early approaches
\cite{Piron1976,Aerts1981,Aerts1982,Aerts1983,Piron1989,Piron1990}, and partly solved by introducing
an orthogonality relation, translating part of the structure of the inverse on
$Q$ to the structure of an orthogonality relation on $\Sigma$: $p, q
\in
\Sigma$, then $p \perp q$ iff there exists a test $\alpha \in  Q$ such that $\alpha$ gives with certainty
`yes' if
$S$ is in state $p$ and
$\widetilde\alpha$ gives with certainty `yes' if $S$ is in state $q$. However
the structure of the inverse was in this way only transferred indirectly to a
structure on ${\cal L}$, by demanding that two properties $a$ and $b$ are
orthogonal iff all states that make $a$ actual are orthogonal to all states
that make $b$ actual. A lot of the operational structure of $\widetilde{\ }: Q
\rightarrow Q$ was lost in this way.

In this article we introduce the structure of the inverse within the more
complete scheme of the state property system and this will lead us to define an
ortho state property system. We also want to study this `inverse' structure for
the closure space that is connected through a categorical equivalence to the
state property system, an equivalence of categories that has shown to be very
fruitful for many other fundamental aspects of quantum axiomatics
\cite{ACVVVS1999,ACVVVS2001,VanSteirteghem2000,VanderVoorde2000,
VanderVoorde2001, ADVV2001,ADVV2002,AD2002}. We also introduce two `weakest'
ortho axioms to make the lattice of properties of our ortho state property
system to be equipped with an orthocomplementation, a necessary structure for
quantum axiomatics
\cite{Piron1976,Aerts1981,Aerts1982,Aerts1983,Piron1989,Piron1990}. Let us
recall some definitions and a theorem. \begin{definition}[Cartan Map] If
$(\Sigma,{\cal L},\xi)$ is a state property system then its Cartan map is the
mapping \mbox{$\kappa : {\cal L} \to {\cal P}(\Sigma)$} defined by $\kappa(a)=
\{p\in \Sigma \ \vert\ a \in \xi(p)\}$. This map has the property that
$\kappa(\wedge_i a_i)=\cap_i \kappa(a_i)$. \end{definition}

\begin{definition} [Closure Space]
\label{def:clos} A closure space $(\Sigma,\mathcal{C})$ consists of a set
$\Sigma$ and a family of subsets \mbox{$\mathcal{C} \subseteq {\cal
P}(X)$}, which are called closed subsets, such that $\emptyset\in\mathcal{C}$
and for $(F_i)_i \in \mathcal{C}$ we have $\cap_iF_i \in \mathcal{C}$.
\end{definition}

\begin{theorem} 
If $(\Sigma,{\cal L},\xi)$ is a state property system then
$(\Sigma,\kappa(\mathcal{L}))$ is a closure space, called the eigenclosure of
$(\Sigma,{\cal L},\xi)$.
Conversely, if $(\Sigma,\mathcal{C})$ is a closure space then
$(\Sigma,\mathcal{C},\bar\xi)$ is a state property system. Here $\mathcal{C}$
is the complete lattice of closed sets, ordered by inclusion and
$\bar\xi:\Sigma\to \mathcal{P}(\mathcal{C}):p\mapsto \{A\in \mathcal{C}|p\in
A\}$.
\end{theorem} 
\noindent For a proof of this theorem we refer to \cite{ACVVVS1999}. 

\section{Ortho State Property Systems}

We are now ready introduce the following
concept of orthogonality:

\begin{definition}[Ortho State Property System] \label{def:orthostatprop}
An ortho state property system $(\Sigma,\mathcal{L},\xi,\orthrel)$ is a state
property system $(\Sigma,\mathcal{L},\xi)$ and a relation $\orthrel$ on
$\mathcal{L}$ such that:\\
\centerline{$\begin{array}{rcl|rcl}
a \orthrel b &\Rightarrow& b \orthrel a &a_i \orthrel b_j\ \forall i, j&\Rightarrow& \wedge_ia_i \orthrel
\wedge_jb_j \\ a \orthrel b &\Rightarrow& a\wedge b = \zero &\zero \orthrel a & &\forall a \in {\cal L}
\end{array}$}
\end{definition}
\noindent The definition of an ortho state property system is inspired by the following: if $a$ and $b$ are
properties and there exist a test $\alpha$ such that $\alpha$ tests $a$ and $\widetilde\alpha$ tests
$b$, then the requirements of definition \ref{def:orthostatprop} follow. A trivial example of a
$\orthrel$ relation is where we would state
$a\orthrel b \Leftrightarrow a \wedge b = \zero$. From now on we will assume to
work with an ortho state property system
$(\Sigma,\mathcal{L},\xi,\orthrel)$, unless explicitly stated otherwise.
We can define the traditional orthogonality relation on the set of states
by means of this relation $\orthrel$.

\begin{proposition}
\label{prop:osps2ortho}
$\orthrel $ induces an orthogonality relation (anti-reflexive, symmetric)
$\perp$ on the set of states $\Sigma$ in the following way: $p \perp q$ if and
only if there are $a, b \in {\cal L}$ such that $a \orthrel b \text{ and } a\in
\xi(p) \text{ and } b \in \xi(q)$ \end{proposition}
\noindent
Now that we have an orthogonality relation on $\Sigma$, it generates the
orthoclosure $(\Sigma,\mathcal{C}_{orth})$ by means of Birkhof's biorthogonal
construction: $\mathcal{C}_{orth}=\{A^{\perp\perp}|A\subset
\Sigma\}$, where $A^\perp=\{p\in \Sigma|\forall q\in A:p\perp q\}$. Conversely
an orthogonality relation on a state property system induced a
$\orthrel$-relation on it's property lattice, as is shown in the following
proposition.

\begin{proposition}
\label{prop:ortho2osps}
If, for a state property system $(\Sigma,\mathcal{L},\xi)$ with an
orthogonality relation $\perp$ on its states, we define $a \orthrel b$
if and only if $p \perp q\ \forall\ p, q \in \Sigma \text{ such that } a \in
\xi(p) \text{ and } b \in \xi(q)$. Then $(\Sigma,\mathcal{L},\xi,\orthrel)$ is
an ortho state property system. \end{proposition}

\begin{proof}
Symmetry of $\orthrel$ is evident. The fact that $\perp$ is antireflexive
implies that whenever $a\orthrel b$, we get that $a\wedge b=\zero$.
Since $\zero\not \in \xi(p)$ for any $p\in \Sigma$ we have that
$p \perp q\ \forall\ p, q \in \Sigma \text{ such that } \zero \in \xi(p) \text{ and
} a \in \xi(q)$ is always true, hence $\zero\orthrel a$ for any $a \in
\mathcal{L}$. Finally $\forall i,j p \perp q\ \forall\ p, q \in \Sigma \text{
such that } a_i \in \xi(p) \text{ and } b_j \in \xi(q)$
implies $p \perp q\ \forall\ p, q \in \Sigma \text{ such
that } \wedge_i a_i \in \xi(p) \text{ and } \wedge_j b \in \xi(q)$
hence we get $a_i \orthrel b_j\ \forall i, j\Rightarrow \wedge_ia_i \orthrel
\wedge_jb_j$.
\end{proof}

\section{Orthocouples and Orthoproperties}
There is another type of orthogonality structure that we can introduce.

\begin{definition}[Orthocouple, Orthoproperty]
\label{def:osps2ocompl}
If $a,b\in \mathcal{L}$ satisfy
\begin{eqnarray*}
b \in \xi(p) &\Leftrightarrow& p \perp q\ \forall\ q\ {\rm such\ that}\ a
\in \xi(q)  \\
a \in \xi(q) &\Leftrightarrow& q \perp p\ \forall\ p\ {\rm such\ that}\ b
\in \xi(p)
\end{eqnarray*}
they form an orthocouple. From this it follows that if $a, b$ and $a, c$ are
orthocouples, then $b = c$. A property $a \in {\cal L}$ which is member of an
orthocouple $a, b$ is called an orthoproperty. For an orthoproperty $a \in
{\cal L}$ we denote the unique property that is defined by it being member of
an orthocouple by $a'$. \end{definition}

\begin{proposition}
If $a, b \in {\cal L}$ are orthoproperties we have $(a')'=a$ and $a<b$ implies
$b'<a'$. \end{proposition}

\begin{proof}
We have $a \in \xi(p) \Leftrightarrow p \perp q\ \forall q$ such
that $a' \in \xi(q) \Leftrightarrow (a')' \in \xi(p)$. This proves that
$(a')' = a$. Suppose that $a < b$ and consider $b' \in \xi(p)$. Then $p
\perp q\ \forall q$ such that $b \in \xi(q)$. Since $a < b$ we also have $p
\perp q\ \forall q$ such that $a \in \xi(q)$. Hence $a' \in \xi(p)$. This
proves that $b' < a'$.
\end{proof}
The relation between the Cartan map $\kappa$ and the $\perp$-relation is
described as in the next propositions.

\begin{proposition}
For an orthoproperty $a \in {\cal L}$ we have $\kappa(a')=\kappa(a)^\perp$ and
$\kappa(a)=\kappa(a)^{\perp\perp}$. \end{proposition}

\begin{proof}
We have $p \in \kappa(a') \Leftrightarrow a' \in \xi(p)
\Leftrightarrow p \perp q\ \forall q$ such that $a \in \xi(q)
\Leftrightarrow p
\perp q\ \forall q$ such that $q \in \kappa(a) \Leftrightarrow p \in
\kappa(a)^\perp$. We remark that for $A
\subset \Sigma$ we have $A^{\perp\perp\perp} = A^\perp$. From this it follows
that $\kappa(a) = \kappa(a')^\perp$. Hence $\kappa(a)^{\perp\perp} =
\kappa(a')^{\perp\perp\perp} = \kappa(a')^\perp = \kappa(a)$.
\end{proof}

\begin{proposition} \label{prop:orthoprop}
A property $a \in {\cal L}$ is an orthoproperty iff $ \kappa(a) =
\kappa(a)^{\perp\perp}$ and there exists $b \in {\cal L}$ such that $\kappa(b)
= \kappa(a)^\perp$ or equivalently iff $\kappa(a) \in {\cal C}_{orth}$ and there exists $b \in {\cal
L}$ such that $\kappa(b) = \kappa(a)^\perp$. In this case $b = a'$.
\end{proposition}

\section{The Ortho Axioms}

This gives us all the material that we need to put forward the first ortho
axiom for an ortho state property system $(\Sigma, {\cal L}, \xi, \orthrel)$.

\begin{axiom}[AO1]\label{AO1}
Axiom Ortho 1 is satisfied if there exists a generating set ${\cal T}$ of
orthoproperties for ${\cal L}$, i.e. $ {\cal L} = \{\wedge_ia_i\ \vert a_i \in
{\cal T}\}$. \end{axiom}
\noindent
The axiom to prolongate the orthocomplementation to the whole of ${\cal L}$
can also easily be put forward now.

\begin{axiom}[AO2]\label{AO2}
Axiom Ortho 2 is satisfied if for $p \in \Sigma$ there exists a property $a_p
\in {\cal L}$ such that $a_p \in \xi(q) \Leftrightarrow q \perp p$. This
implies the uniqueness of $a_p$. \end{axiom}

\begin{definition} [Orthocomplementation]
Suppose that we have a state property system $(\Sigma, {\cal L}, \xi)$. A
function $': {\cal L} \rightarrow {\cal L}$,
such that for $a , b \in {\cal }$ we have:
$$\begin{array}{rcl|rcl}
(a')' &=& a &a \wedge a' &=& \zero\\
a < b &\Rightarrow& b' < a'  &a \vee a' &=& \one
\end{array}$$
is called an orthocomplementation of ${\cal L}$.
\end{definition}

\begin{theorem}
We have:
\begin{itemize}
\item[(A)] If an ortho state property system
$(\Sigma,\mathcal{L},\xi,\orthrel)$ satisfies AO1 and AO2 then it induces an
orthocomplementation $':\mathcal{L}\to \mathcal{L}$ on the state property
system $(\Sigma,\mathcal{L},\xi)$. Here $a'$ is defined as the unique member of
$\mathcal{L}$ for which $a,a'$ forms an orthocouple in
$(\Sigma,\mathcal{L},\xi,\orthrel)$.

\item[(B)] If a state property system $(\Sigma,\mathcal{L},\xi)$ has an
orthocomplementation $':\mathcal{L}\to \mathcal{L}$ then
$(\Sigma,\mathcal{L},\xi,\orthrel)$ is an ortho state property system
satisfying AO1 and AO2, where $\orthrel$ is defined by
$a\orthrel b \Leftrightarrow b<a'$.

\end{itemize}
\end{theorem}

\begin{proof}

(A): We prove that for $a\in\mathcal{L}$ there exists a unique property $a'$ such that:
$a' \in \xi(p) \Leftrightarrow p \perp q\ \forall q \in \Sigma\ {\rm such\
that}\ a \in \xi(q)$.
The fact that for each $p\in \Sigma$ AO2 gives us a unique $a_p\in
\mathcal{L}$, allows us to define $a'=\wedge_{a\in \xi(p)}a_p$.
Suppose that $a'\in \xi(r)$ for some $r\in\Sigma$ then obviously for a
$p\in \Sigma$ such that $a\in \xi(p)$ we get $a'<a_p$. Hence $a_p\in \xi(r)$,
for each $p\in\Sigma$ such that $a\in \xi(p)$. Therefore $r\perp p$ for each
$p\in\Sigma$ such that $a\in \xi(p)$. Conversely, if $r\perp q$ for every $q\in
\Sigma$ such that $a\in\xi(q)$ we know from AO2 that $a_q\in \xi(r)$ for every
such $q$. Hence $a'=\wedge_{a\in \xi(q)}a_q\in \xi(r)$. By the last two
results we have that our $a'$ is the same as $a'$ defined by Definition
\ref{def:osps2ocompl}. There remains to prove that $'$ is indeed an
orthocomplementation. For $a\in \mathcal{L}$ we know that there exists
orthoproperties $a_i$ such that $a=\wedge_i a_i$, for which $(a_i')'=a_i$. Thus
$(a')'\in \xi(r)$ this is equivalent to
\begin{eqnarray*}
&&r \perp q \text{ for every $q\in \Sigma$ such that $a'\in \xi(q)$}\\
&\Leftrightarrow&r \perp q \text{ for every $q\in \Sigma$ such that $q\perp p$ for every
$p\in \Sigma$ such that $a\in\xi(p)$}\\
&\Leftrightarrow&\forall i:r \perp q \text{ for every $q\in \Sigma$ such that $q\perp p$ for every
$p\in \Sigma$ such that $a_i\in\xi(p)$}\\
&\Leftrightarrow&\forall i:r \perp q \text{ for every $q\in \Sigma$ such that $a_i'\in
\xi(q)$}\\
&\Leftrightarrow&\forall i:(a_i')'=a_i\in \xi(r)\\
&\Leftrightarrow&a\in \xi(r)
\end{eqnarray*}
Hence $(a')'=a$. If $a<b$ and $b'\in \xi(r)$ we have that $r\perp q$ for every
$q\in \Sigma$ such that $b\in \xi(q)$, but $a\in \xi(q)$ implies $b\in \xi(q)$
hence $r\perp q$for every $q\in \Sigma$ such that $a\in \xi(q)$, so $a'\in
\xi(r)$ and thus $b'<a'$. Finally, if $\zero\not=a\wedge a'\in\xi(r)$ then
$a,a'\in \xi(r)$ but this implies $r\perp r$ which is impossible, so $a \wedge
a'=\zero$. Analogously $a\vee a'=\one$.

(B): We only give the proof of the last condition on $\orthrel$, the
others are easy verifications. Let $a_i\orthrel b_j$ for every $i,j$, then
by definition of $\orthrel$ we get $b_j<a_i'$, so $b_j<\wedge_i a_i'$. We also
have $\wedge_i a_i<a_i$, hence $a_i'<(\wedge_i a_i)'$ so that $\wedge_i
a_i'<(\wedge_i a_i)'$. Therefore $b_j<(\wedge_i a_i)'$ and thus $\wedge_j
b_j<(\wedge_i a_i)'$. Finally we conclude that $\wedge_i a_i \orthrel \wedge_j
b_j$. In order to prove AO1 we will prove that every pair $a,a'$ is an
orthocouple, i.e.:\\
\centerline{$\begin{array}{ccc}
a\in \xi(q) &\Leftrightarrow& q \perp p \ \forall \ p \text{ such
that } a'\in \xi(p)\\
a'\in \xi(p) &\Leftrightarrow& p \perp q \ \forall \ q \text{ such that } a\in
\xi(q)
\end{array}$}\\
where $\perp$ is given by
$p\perp q\Leftrightarrow \exists  a,b\in\mathcal{L}:a\orthrel b, a\in \xi(p), b\in
\xi(q)$.
We prove the first statement, the second is completely analogous. Let $a\in
\xi(q)$ and $p$ such that $a'\in \xi(p)$ then obviously there $a\orthrel a'$,
hence $p\perp q$. Conversely, suppose $p\perp q$ for each $p$ such that $a'\in
\xi(p)$. So for such a $p$ there are $\tilde a\in \xi(q),\tilde b\in \xi(p)$
for which $\tilde a \orthrel \tilde b$, so $\tilde b<\tilde a'$. Since $\tilde
b\in \xi(p)$, we know that $\tilde a'\in \xi(p)$. We have:\\
\centerline{$\forall p \text{ such that } a'\in \xi(p):\exists \tilde a_p\in
\xi(q): \tilde a_p'\in\xi(p)$}
We now define $c=\wedge_{a'\in\xi(p)} \tilde a_p$. Since
$c<\tilde a_p$, we have that $\tilde a_p'<c'$, hence $c'\in \xi(p)$, for each
$p$ such that $a'\in\xi(p)$. This means that $a'<c'$. Using the
orthocomplementation we get $c<a$, but since $c\in \xi(q)$, $a\in \xi(q)$.
Hence AO1 holds. From the above we also see that the orthocomplementation
induced by the ortho state property system $(\Sigma,\mathcal{L},\xi,\orthrel)$,
is the same as the given one, since $a,a'$ always form an orthocouple. In order
to prove AO2 we choose an $p\in \Sigma$ and consider $a_p=(\wedge \xi(p))'$. If
$a_p\in \xi(q)$ then for $\tilde a=\wedge\xi(p)\in\xi(p)$ and $\tilde
b=a_p\in\xi(q)$ we have that $\tilde b<\tilde a'$, hence $\tilde a \orthrel
\tilde b$ and thus $p\perp q$. Conversely, if $p \perp q$ then there are
$\tilde a\in \xi(p),\tilde b\in \xi(q)$ for which $\tilde b<\tilde a'$. Since
$a_p'=\wedge \xi(p)<\tilde a$ we know that $\tilde a'<a_p$, so $\tilde b<a_p$
which implies $a_p\in \xi(q)$. Hence AO2 also holds and we have proven the
theorem. \end{proof}
From the proof of this theorem one has the following corollary which we shall
need further on.

\begin{corollary}
Take an ortho state property system $(\Sigma, {\cal L}, \xi, \orthrel)$ for
which AO1 and AO2 are satisfied. Let $':\mathcal{L}\to \mathcal{L}$ be the
orthocomplementation described in the theorem, then $a'=\wedge_{a\in
\xi(p)}a_p$ where the $a_p$ are given by AO2. Moreover every property of
$(\Sigma, {\cal L}, \xi, \orthrel)$ is an orthoproperty. \end{corollary}

\section{Eigenclosure and Orthoclosure}

The previous Theorem describes the link between an ortho state property
system $(\Sigma,\mathcal{L},\xi,\orthrel)$ and an orthocomplementation
$':\mathcal{L}\to\mathcal{L}$. In what follows we'll turn our attention towards
the associated closure spaces: the eigenclosure and the orthoclosure.

\begin{theorem}
\label{thm1}
Consider an ortho state property system $(\Sigma,\mathcal{L},\xi,\orthrel)$, then:
$$\kappa(\mathcal{L})=\mathcal{C}_{orth} \Leftrightarrow \text{ AO1 and AO2 }$$
\end{theorem}

\begin{proof}
Let $A$ be closed in $(\Sigma,\mathcal{C}_{orth})$, i.e. $A=A^{\perp\perp}$. By
AO2 we know that $\forall p\in A:\exists a_p\in \mathcal{L}: a_p\in \xi(q)\Leftrightarrow
q \perp p$. We make $a=\wedge \{a_p|p\in A\}$ and define $A^*=\kappa(a)$.
Then $q\in A^*$ is equivalent to $q\in \kappa(\wedge \{a_p|p\in A\})=\cap_{p\in
A}\kappa(a_p)$. So for any $p\in A$ one has that $q\in \kappa(a_p)$, which
means that $a_p\in \xi(q)$. Using (AO2) we obtain that $q\in A^*$ is
equivalent to $p\perp q$ for every $p\in A$, so $q\in A^{\perp\perp}$
Hence $A^*=A^{\perp\perp}=\kappa(a)$ which is closed in
$(\Sigma,\kappa(\mathcal{L}))$. Conversely, if $A$ is closed in
$(\Sigma,\kappa(\mathcal{L}))$, there is an $a\in \mathcal{L}$ such that
$A=\kappa(a)$. Since $a$ is an orthoproperty we know that
$\kappa(a)=\kappa(a)^{\perp\perp}$ so $A=A^{\perp\perp}$ is closed in
$(\Sigma,\mathcal{C}_{orth})$.

Let $p\in \Sigma$. $\{p\}^\perp\in \mathcal{C}_{orth}$ since
$\{p\}^{\perp\perp\perp}=\{p\}^{\perp}$ since
$\kappa(\mathcal{L})=\mathcal{C}_{orth}$ there is a property $a$
such that $\kappa(a)=\{p\}^{\perp}$. For this $a$ we have the following chain
of equivalences $a\in \xi(q) \Leftrightarrow q\in \kappa(a) \Leftrightarrow
q\in \{p\}^{\perp} \Leftrightarrow q \perp p$.
So for any $p$ there is an $a=a_p$ such that $a\in \xi(q)\Leftrightarrow q \perp p$, hence
AO2 follows. Clearly $\mathcal{T}=\mathcal{L}$ is a generating set. Take $a\in
\mathcal{L}$ then $\kappa(a)\in \kappa(\mathcal{L})=\mathcal{C}_{orth}$, so
$\kappa(a)=\kappa^{\perp\perp}$. By Definition of $\mathcal{C}_{orth}$ we know
that $\kappa(a)^\perp\in \mathcal{C}_{orth}=\kappa(\mathcal{L})$, so there is
a $b\in \mathcal{L}$ such that $\kappa(b)=\kappa(a)^\perp$. By proposition
\ref{prop:orthoprop} we see that $a$ is an orthoproperty, hence AO1 also holds.
\end{proof}

\begin{theorem}
We have:
\begin{itemize}
\item[(C)] If an ortho state property system
$(\Sigma,\mathcal{L},\xi,\orthrel)$ satisfies AO1 and AO2 then the closure
space $(\Sigma,\kappa(\mathcal{L}))$ is induced by the underlying
$\perp$-relation of Proposition \ref{prop:osps2ortho}.

\item[(D)] If a closure space $(\Sigma,\mathcal{C})$ is induced by a
$\perp$-relation then $(\Sigma,\mathcal{C},\bar\xi,\orthrel)$ is an ortho state
property system satisfying AO1 and AO2, where $\orthrel$ is as in Proposition
\ref{prop:ortho2osps}.
\end{itemize}
\end{theorem}

\begin{proof}

(C): Follows from the previous Theorem \ref{thm1} and Proposition
\ref{prop:osps2ortho}.

(D): Let $\mathcal{C}$ be induced by an orthogonality relation $\perp$.
From Proposition \ref{prop:ortho2osps} we know that
$(\Sigma,\mathcal{C},\bar\xi,\orthrel)$ is an ortho state property system. Let
us denote by $\perp^*$ the underlying $\perp$-relation, i.e. $p\perp^* q
\Leftrightarrow \exists A,B\in \mathcal{C}:A\orthrel B, p\in A, q\in B$.
Suppose that $p\perp^* q$, then there are $A,B\in \mathcal{C}$ with $p\in A$
and $q\in B$ such that $A\orthrel B$. Hence for any $\tilde p\in A$ and $\tilde
q\in B$ we have that $\tilde p\perp \tilde q$, so since $p\in A$ and $q\in B$
we get that $p\perp q$. Conversely, if $p\perp q$, then we choose
$A=\{p\}^{\perp\perp}$ and $B=\{q\}^{\perp\perp}$. Obviously $p\in A$ and $q\in
B$, moreover if $\tilde p\in A$ and $\tilde q\in B$ then $\tilde p\perp r$ for
any $r$ such that $r\perp p$. In particular for $r=q$ we have $\tilde p\perp
q$, hence $\tilde p\in\{q\}^\perp$, so $\tilde q\perp \tilde p$. Thus
$A\orthrel B$. Finally we have: $\exists A,B\in \mathcal{C}:A\orthrel B, p\in
A, q\in B$, so we conclude that $p\perp^* q$. By the equivalence between
closure spaces and state property systems we know that
$\kappa(\mathcal{C})=\mathcal{C}=\mathcal{C}_{orth}$, hence by Theorem
\ref{thm1} we have that AO1 and AO2 are fulfilled. \end{proof}
With the above results (A),(B),(C) and (D) we consider the following scheme.
We start with an ortho state property system
$(\Sigma,\mathcal{L},\xi,\orthrel)$ satisfying AO1 and AO2.
First we use (A) to get a state property system $(\Sigma, \mathcal{L}, \xi)$
and an orthocomplementation $':\mathcal{L}\to \mathcal{L}$. Applying (B) we get
a new ortho state property system $(\Sigma, \mathcal{L}, \xi ,\orthrel^*)$,
which satisfies AO1 and AO2. On the other hand we can apply (C) to the ortho
state property system $(\Sigma,\mathcal{L},\xi,\orthrel)$, hence we get a
closure space $(\Sigma, \kappa(\mathcal{L}))$ where
$\kappa(\mathcal{L})=\mathcal{C}_{orth}$ is induced by the orthogonality
relation $\perp$ of $(\Sigma,\mathcal{L},\xi,\orthrel)$. Using (D) we get again
an ortho state property system $(\Sigma, \kappa(\mathcal{L}), \bar\xi
,\orthrel^{**})$, satisfying AO1 and AO2. We now ask ourselves what the
relation is between those three ortho state property systems. First we note
that by the general equivalence between state property systems and closure
spaces $(\Sigma,\mathcal{L},\xi)$ and $(\Sigma,\kappa(\mathcal{L}),\bar \xi)$
can be considered as being the same (up to isomorphism). The relation between
$\orthrel^*$ and $\orthrel^{**}$ is given in the following theorem.

\begin{theorem}
With the above notations, we have
$\kappa(a)\orthrel^{**}\kappa(b)\Leftrightarrow a\orthrel^*b$.
\end{theorem}

\begin{proof}
$\kappa(a)\orthrel^{**}\kappa(b)$ is equivalent to $p\perp q, \forall p,q
\text{ such that } a\in\xi(p) \text{ and } b\in\xi(q)$. Thus, for any $p$ such
that $a\in \xi(p)$, one has that $b\in\xi(q)$ implies $p\perp q$. By means of
(AO2) it also implies $b<a_p$, hence $\kappa(a)\orthrel^{**}\kappa(b)$ is
equivalent to $b<\wedge_{a\in\xi(p)}a_p=a'$, which means that $a\orthrel^*b$.
\end{proof}

\noindent From this we know that $(\Sigma, \mathcal{L}, \xi ,\orthrel^*)$ and
$(\Sigma, \kappa(\mathcal{L}), \bar\xi ,\orthrel^{**})$ are essentially the
same. In order to compare $(\Sigma, \mathcal{L}, \xi ,\orthrel^*)$ with the
original ortho state property system $(\Sigma,\mathcal{L},\xi,\orthrel)$ we
need one more proposition.

\begin{proposition}
For any ortho state property system $(\Sigma,\mathcal{L},\xi,\orthrel)$
satisfying AO1 and AO2 we have $a\orthrel b\Rightarrow b<a'$.
\end{proposition}

\begin{proof}
$a\orthrel b$ implies $p\perp q, \forall p,q \text{ such that }
a\in\xi(p) \text{ and } b\in\xi(q)$. With the same reasoning as in the previous
proof one finds that it also implies $b<a'$. \end{proof}

\begin{theorem}
$\orthrel^*$ is the largest relation such that
$(\Sigma,\mathcal{L},\xi,\orthrel^*)$ is an ortho state property system,
satisfying AO1 and AO2, with the same orthocomplementation
$':\mathcal{L}\to\mathcal{L}$ as $(\Sigma,\mathcal{L},\xi,\orthrel)$.
\end{theorem}

\begin{proof}
Consider a relation $\tilde{\orthrel}$ such that
$(\Sigma,\mathcal{L},\xi,\tilde{\orthrel})$ is an ortho state property system,
satisfying AO1 and AO2, with the same orthocomplementation
$':\mathcal{L}\to\mathcal{L}$ as $(\Sigma,\mathcal{L},\xi,\orthrel)$.
By the previous proposition we have $a\tilde{\orthrel}b$ which
implies $b<a'$. Therefore $a\orthrel^*b$ and thus $\tilde{\orthrel}\subset
\orthrel^*$. \end{proof}

\noindent To conclude we give an example showing that $\orthrel^*$ can be
strictly larger than $\orthrel$.

\begin{example}
Consider a set of states $\Sigma=\{p,q,r,s,t,u\}$ and the property lattice
$\mathcal{L}$ (see figure \ref{fig1}), with top $\one=10$ and bottom $\zero=1$.

\begin{figure*}[ht]
\centering
\includegraphics[width=3cm,height=3cm]{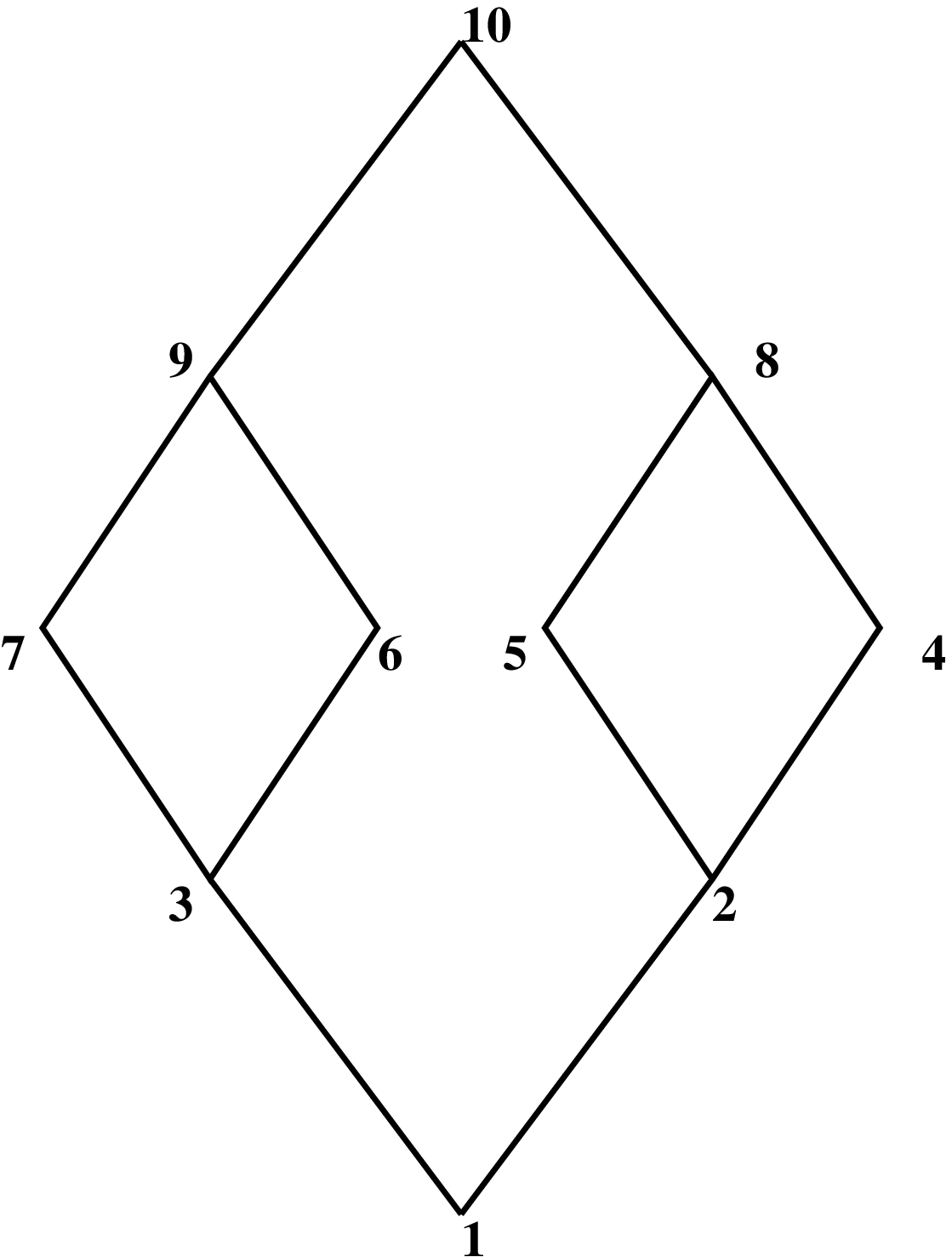}
\includegraphics[width=4cm,height=3cm]{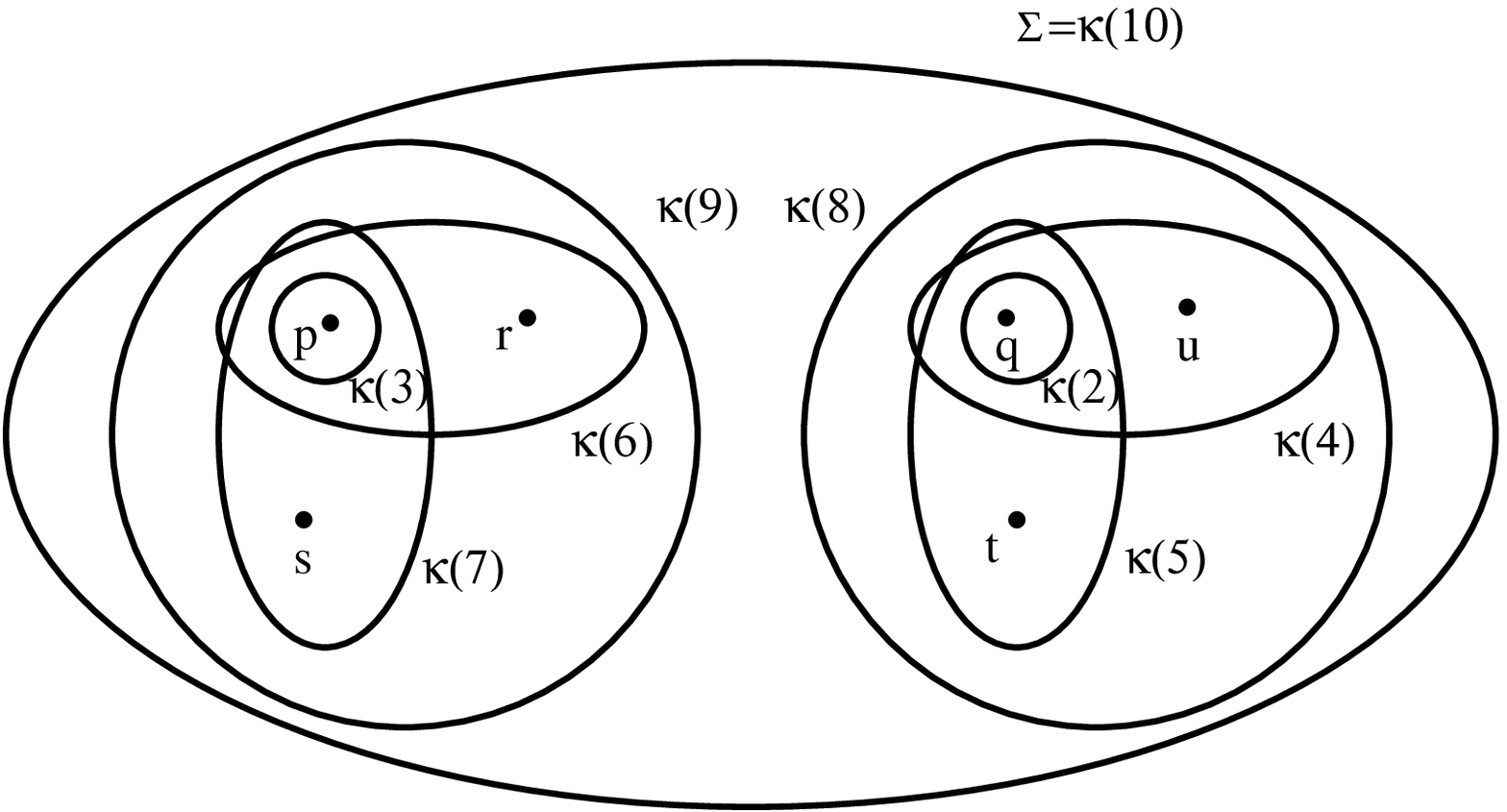}
\caption{The lattice $\mathcal{L}$ and the closure space
$\kappa(\mathcal{L})$.} \label{fig1}
\end{figure*}
\noindent
We define the map $\xi$ by $\xi(p)=\{3,6,7,9,10\}, \xi(q)=\{2,4,5,8,10\},
\xi(r)=\{6,9,10\}, \xi(s)=\{5,8,10\}, \xi(t)=\{7,9,10\}, \xi (u)=\{4,8,10\}$.
In this way $(\Sigma,\mathcal{L},\xi)$ is a state property system.
We endow it with the following relation:
$$\orthrel=\{(i,1),(1,i)|1\leq i \leq 10\} \cup \{(7,5),(5,7),(4,6),(6,4)\}$$
Hence we get an ortho state property system
$(\Sigma,\mathcal{L},\xi,\orthrel)$. Since
$\mathcal{C}_{orth}=\kappa(\mathcal{L})$ we have that both AO1 and AO2 are
satisfied. We can now consider $\orthrel^*$. Since $\kappa(2)=\{q\}$,
$\kappa(3)=\{p\}$ and $p \perp q$, we have that
$\kappa(2)\orthrel^{**}\kappa(3)$, hence $2 \orthrel^* 3$. So $\orthrel^*$ is
strictly larger than $\orthrel$. In fact it is given by: \begin{eqnarray*}
\orthrel^*=\orthrel &\cup& \{(3, 2), (2, 3), (4, 3), (3, 4), (3, 5), (2, 7),\\
&&(2, 6), (7, 2), (6, 2), (2, 9), (9, 2), (5, 3),(3, 8), (8, 3)\}
\end{eqnarray*}

\end{example}


\end{document}